\documentclass[10pt,a4paper,twocolumn]{article}
\usepackage{graphicx} 
\usepackage{siunitx} 
\usepackage{amsmath} 
\usepackage{amstext} 
\usepackage{amssymb}
\usepackage[top=25mm,bottom=37mm,left=20mm,right=20mm,columnsep=10mm]{geometry} 
\usepackage{color} 
\definecolor{myblu}{rgb}{0.1,0.1,0.5}
\usepackage{hyperref} 
\hypersetup{colorlinks=true,urlcolor=blue,linkcolor=black,citecolor=black} 
\usepackage{sectsty,textcase} 
\sectionfont{\large\MakeTextUppercase} 
\usepackage{secdot} 
\usepackage{tikz}
\usepackage{subcaption}

\usepackage{xcolor}

\definecolor{mplblue}{RGB}{0,114,189}
\definecolor{mplred}{RGB}{213,94,0}
\definecolor{mplgreen}{RGB}{86,177,76}

\newcommand{\ccontline}[1]{$\vcenter{\hbox{{\protect\tikz{\protect\draw[#1,-,line width=1.5pt] (0,0) -- (0.4,0);}}}}$}
\newcommand{\cdashedline}[1]{$\vcenter{\hbox{{\protect\tikz{\protect\draw[#1,dashed,line width=1.5pt] (0,0) -- (0.4,0);}}}}$}

\newcommand{\contline}{$\vcenter{\hbox{{\protect\tikz{\protect\draw[-,line width=1.5pt] (0,0) -- (0.4,0);}}}}$}
\newcommand{\dotline}{$\vcenter{\hbox{{\protect\tikz{\protect\draw[dotted,line width=1.5pt] (0,0) -- (0.4,0);}}}}$}
\newcommand{\dashline}{$\vcenter{\hbox{{\protect\tikz{\protect\draw[dashed,line width=1.5pt] (0,0) -- (0.4,0);}}}}$}

\begin{document}
\title{\vspace{-18mm}
\begin{minipage}{\linewidth}
\hspace{5mm}\raisebox{-50pt}{\includegraphics[width=.23\textwidth]{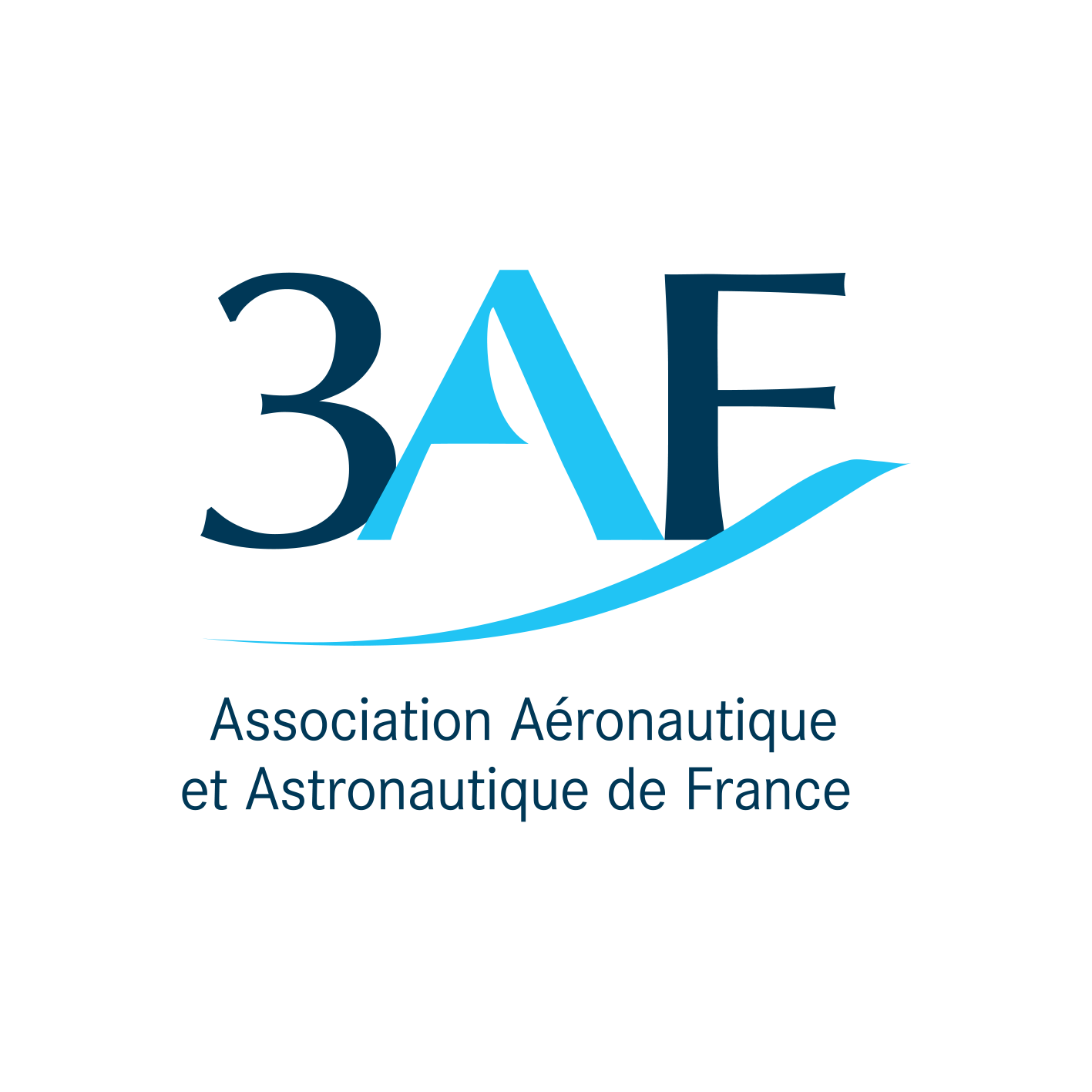}}\hspace{4mm}
\textcolor{myblu}{\textbf{\textit{\small\begin{tabular}{l r}
57$^\text{th}$ 3AF International Conference\\ on Applied Aerodynamics\\ 29 --- 31 March 2023, Bordeaux -- France
\end{tabular}}}}
\hspace{1mm}\textbf{\small AERO2023-64-CAILLAUD}
\end{minipage}\\\vspace{10mm}
    \textbf{\Large Global stability analysis of a hypersonic cone-cylinder-flare geometry}}
    \author{\textbf{\normalsize Clément Caillaud$^\text{(1)}$, 
    Mathieu Lugrin$^\text{(2)}$,
    Sébastien Esquieu$^\text{(3)}$,
    Cédric Content$^\text{(4)}$
    }
    \\{\normalsize\itshape
    $^\text{(1)}$CEA-CESTA, 15 Avenue des Sablières, Le Barp, France, clement.caillaud@cea.fr}
    \\{\normalsize\itshape
    $^\text{(2)}$DAAA, ONERA, Université Paris Saclay F-92190 Meudon - France, mathieu.lugrin@onera.fr}
    \\{\normalsize\itshape
    $^\text{(3)}$CEA-CESTA, 15 Avenue des Sablières, Le Barp, France, sebastien.esquieu@cea.fr}
    \\{\normalsize\itshape
    $^\text{(2)}$DAAA, ONERA, Université Paris Saclay F-92322 Châtillon - France, cedric.content@onera.fr}
    }
\date{}

\maketitle
\begin{abstract}
    Characterizing the boundary layer transition to turbulence around realistic hypersonic vehicles is a challenging task due to the numerous parameters that affect the process. To address this challenge, the cone-cylinder-flare (CCF) geometry has been designed to provide a flow topology that captures various transition mechanisms observed on reentry objects \cite{esquieuFlowStabilityAnalysis2019}, such as absolute and convective instabilities, which are dependent on the free stream conditions. In this study, a global linear stability analysis is performed on the CCF model at $M_\infty=6.0$ to investigate and map the dominant instabilities at wind tunnel flow conditions. We examine optimal responses and forcings computed using resolvent analysis, as well as global modes originating from the recirculation bubble at the cylinder flare junction. The effects of bluntness are assessed through analyses of both blunt and sharp configurations. Our results shed light on the linear flow mechanisms that promote the transition to turbulence around such hypersonic objects.
\end{abstract}

\section{Introduction}
\begin{figure*}
    \includegraphics*[width=0.9\textwidth]{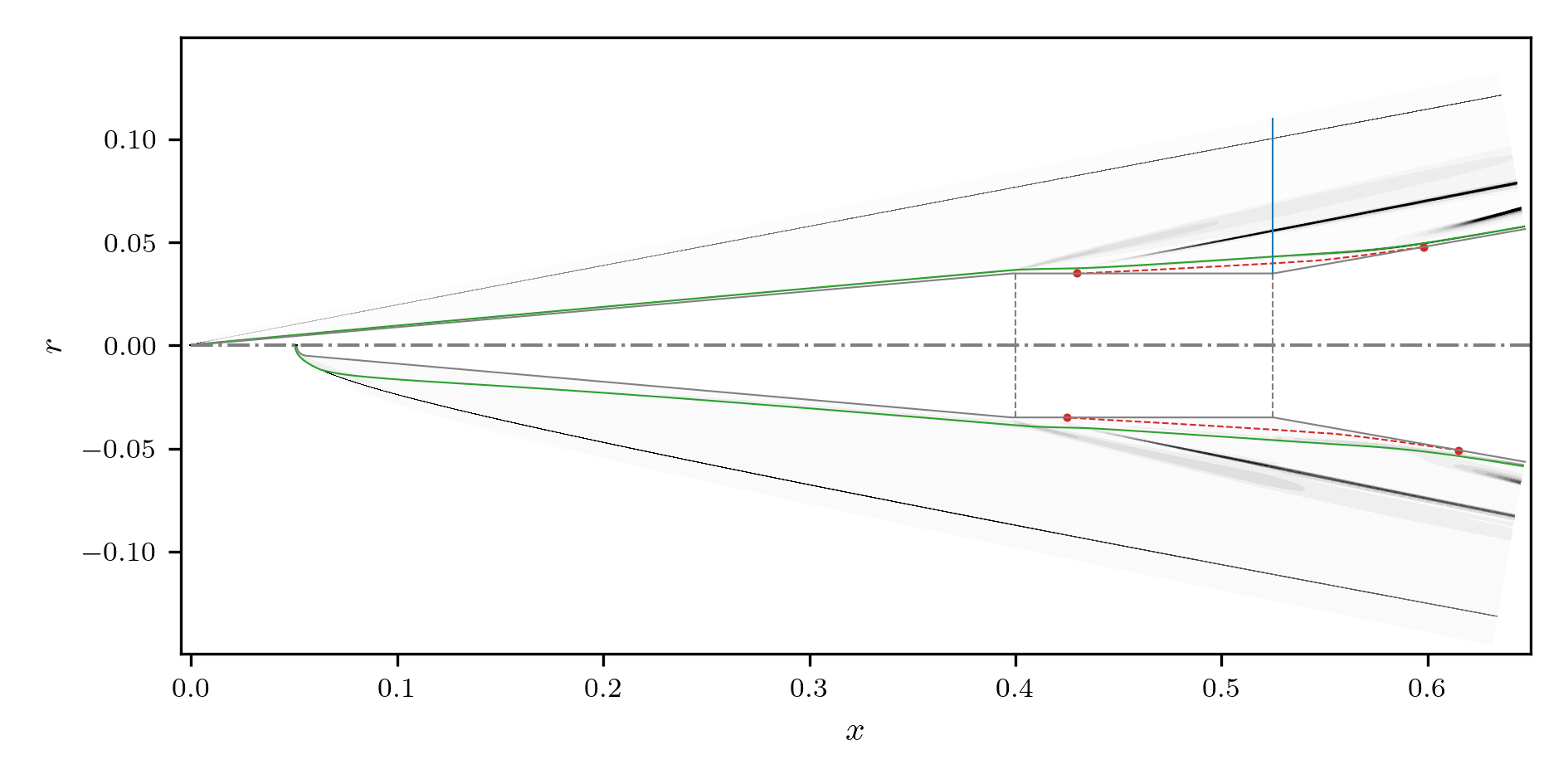}
    \caption{Baseflow for the  $R_n=0.1\text{mm}$ (top) \&  $R_n=5\text{mm}$ (bottom) CCF10 geometries. Greyscale : normalised numerical shadowgraphy with \ccontline{mplgreen}  : entropy layer ; \cdashedline{mplred} : recirculation bubble with separation and reattachment points (\textcolor{mplred}{$\bullet$}) ; \ccontline{mplblue} : line along which the profiles of Fig. \ref{fig:ccf12-valid} are extracted.}
    \label{fig:ccf12-bf}
\end{figure*}

The current CCF geometry consists in a $5^\circ$ sharp or blunt cone, followed by a cylindrical section which terminates to a flare with an angle of $\theta_f=10^\circ$ (CCF10). The geometry is depicted in Fig. \ref{fig:ccf12-bf} and can be divided into two primary regions: the cone section, which provides a canonical convectively unstable boundary layer that supports first and second mode waves, and the cylinder-flare section, which includes a separation bubble generated by the adverse pressure gradient caused by the shock-wave at the cylinder-flare junction. This flow topology is similar to that described in previous works by \cite{lugrinTransitionScenarioHypersonic2021,dwivediObliqueTransitionHypersonic2022}. In this second region, the separation bubble has been found to be globally unstable for flare angles $\theta_f \geq 8^\circ$ and Reynolds numbers above $Re=11.5\times10^6$ \cite{liNonlinearEvolutionInstabilities2022}. At this critical Reynolds number, convectively unstable modes such as first-mode and second mode waves can also exist along the recirculation bubble and/or the reattachment point. For cases where either the convective, the global instabilities or a combination of both, lead to transition, the non-linear terms will induce a strong modification of the bubble. Therefore making the baseflow  stability analysis irrelevant and leading to a mean flow analysis \cite{marxenEffectSmallamplitudeConvective2011,lugrinTransitionScenarioHypersonic2021}. However, all the cases treated in this article stay fully laminar in the whole domain and thus base flow stability analysis is justified.

For these flare angles, the previous experimental campaigns in quiet and noisy tunnels, associated with numerical investigations using PSE, revealed that both first and second-mode waves are amplified along the sharp cone region. On the cylinder-flare region, second mode instabilities were not found to be growing until the bubble reattachment whereas first-mode waves were seen to be continuously amplified along the bubble \cite{benitezInstabilityMeasurementsAxisymmetric2020,paredesBoundaryLayerInstabilitiesCone2022}. Additionally, infrared thermography revealed the presence of steady streaks originating at the reattachment and amplifying along the flare at an azimuthal wavenumber of $m=36$. The origin of these streaks remain poorly understood but is suggested to be the byproduct of non-linear interaction in the bubble \cite{liNonlinearEvolutionInstabilities2022}.

Considering these previous results, we employ a global linear stability framework to explore the transition dynamics of the CCF10 object. Our aims are twofold: first, characterise the base flow and identify the instabilities present in the separation bubble, and second, analyse the forced response of the linear dynamics through resolvent analysis. To achieve these goals, we organise this paper as follows: in Section 2 describes the numerical stability tools and the base flow, along with a validation. In Section 3 are discussed the global modes found in the bubble, and the Section 4 investigates the forced response of the linear dynamics. Finally, Section 5 draw conclusions and discuss the future directions of this research.

\section{Linear Stability} \label{sec:lin-framework}
The analysis is supported by the stability toolbox BROADCAST \cite{poulainBROADCASTHighorderCompressible2023}. It uses high order finite-volume schemes and algorithmic differentiation to compute fixed points of the compressible Navier-Stokes equations and extract the associated direct and adjoint global linear operators, along with its derivatives up to arbitrary order if required. Using this toolbox, the  methodology is introduced in the next paragraphs. 
For further details about the framework presented in this section, the reader is referred to the comprehensive review of \cite{sippDynamicsControlGlobal2010}.

Considering a domain in cylindrical coordinates $(x,r,\theta)$ discretised in a structured fashion with $N_p$ points. The flow dynamics around the CCF geometry are governed by the following discrete non-linear dynamical system,
\begin{align}
    \frac{\partial \pmb q}{\partial t} = \mathbf{R}(\pmb q) + \pmb{f_e}, \label{eq:nl-system}
\end{align}
where $\pmb q$ is the conservative state vector of $N_v$ variables, $\pmb R$ is the compressible Navier-Stokes operator and $\pmb{f_e}$ is a harmonic exogenous forcing of small amplitude. Supposing the existence of a fixed-point $\pmb q_0$ such that $\pmb{R}(\pmb q_0)=0$, the non-linear system can be linearised around this steady state in order to retrieve the linear dynamics of small disturbances $\pmb q'$, with $\pmb q = \pmb q_0 + \epsilon\pmb q',\ \epsilon \ll 1$. The linear disturbances equations reads,
\begin{align}
    \frac{\partial \pmb q'}{\partial t} = \mathbf{L}\pmb q' + \pmb{f_e}, 
    \quad \text{with } \mathbf{L} = \left.\frac{\partial \pmb{R}(\pmb q)}{\partial \pmb q}\right|_{\pmb q_0}, \label{eq:l-system}
\end{align}
with $\mathbf{L}$, the Jacobian of the non linear operator around $\pmb q_0$. To analyse the disturbances dynamics, we make the hypothesis of a zero-angle of attack, and consider the problem to be 2D axisymmetric. Hence the disturbances and forcing vectors are expressed in terms of their frequency $\omega$ and azimuthal Fourier modes $m$, resulting in the harmonic ansatzs,
\begin{align}
    \pmb q'(x,r,\theta,t) = \hat{\pmb q}(x,r) e^{i (m \theta + \omega t)}, \\ 
    \pmb f_e(x,r,\theta,t) = \hat{\pmb f}(x,r) e^{i (m \theta + \omega t)}. 
\end{align}

Using these vectors, the linear dynamics of Eq. \ref{eq:l-system} will be studied in two ways. First, by considering the autonomous system, i.e $\pmb{f_e} = 0$. In Fourier space, the global stability of the baseflow at a given frequency and wavenumber $m$ (with this formalism, the Jacobian operator becomes dependent on $m$) can be studied by solving the eigenvalue problem,
\begin{equation}
    \mathbf{L}(m)\hat{\pmb q} =  i\omega \hat{\pmb q}, \quad \omega \in \mathbb{C}, \label{eq:EVP}
\end{equation}
with the growth-rate of an eigenfunction given by $\Re(\omega)$ and its frequency by $\Im(\omega)$. 

Next, the stability of the forced system, i.e $\pmb{f_e} \neq 0$, can be studied. Due to the non-normality of the jacobian operator $\mathbf{L}$, the exogenous forcing can trigger the non-modal amplification of responses even if the system is stable \cite{schmidNonmodalStabilityTheory2007}. These non-normal mechanisms are given by the resolvent operator defined as $\pmb{\mathcal{R}}=(i\omega\mathbf{I} - \mathbf{L})^{-1}$. For noise-amplifier flows, this operator yields an input-output relation at a given frequency and wavenumber between forcings and responses over the baseflow, such that,
\begin{align}
    \hat{\pmb q} = \pmb{\mathcal{R}}(\omega,m)\hat{\pmb f}, \quad \omega \in \mathbb{R}. \label{eq:i-o}
\end{align}

To find the most amplified convective instabilities at a pair $(\omega,m)$, an optimal decomposition of the resolvent matrix can be computed and provides an orthogonal basis of optimal forcings and responses ranked by energy. Starting from Eq. \ref{eq:i-o}, such functions can be found by solving an optimisation problem. Using the discrete norms $||\hat{\pmb q}||_E = \hat{\pmb q}^* \mathbf{W}_E \hat{\pmb q}$ and $||\hat{\pmb f}||_{F}=\hat{\pmb f}^*\mathbf{P}^*\mathbf{W}_f \mathbf{P} \hat{\pmb f}$, where $\mathbf{W}_E$ is the Chu energy weight matrix \cite{chuEnergyTransferSmall1965} and $\mathbf{W}_F$ is taken as the identity along with a restriction matrix $\mathbf{P}$ used to impose the forcing on specific regions or variables, a Rayleigh quotient can be obtained,
 \begin{align}
    \mu_0^2 = \sup_{\hat{\pmb f}} \frac{||\hat{\pmb q}||_E }{||\hat{\pmb f}||_{F}} = \sup_{\hat{\pmb f}} \frac{||\mathcal{R} \hat{\pmb f}||_E }{||\hat{\pmb f}||_{F}}. \label{eq:gain}
\end{align}
The optimal forcing that satisfies equation \ref{eq:gain} may then be found by solving the eigenvalue problem :
\begin{align}
    \mathbf{P}^* \pmb{\mathcal{R}}^* \mathbf{W}_E \pmb{\mathcal{R}} \mathbf{P} \hat{\pmb f_i} &= \mu_i^2 \mathbf{W}_f \hat{\pmb f_i}. \label{eq:resolvent-evp}
\end{align}
%
Where $(\bullet)^*$ is the hermitian transpose and the eigenvalues $\mu^2_0>...>\mu^2_i>\mu^2_{i+1}>...$ of Eq. \ref{eq:resolvent-evp} are the optimal gains ranked by energy and their associated eigenfunctions $\hat{\pmb f}_i^{opt}$. Solving Eq. \ref{eq:resolvent-evp} for various values of $(\omega, m)$ allows to map the system resonance peaks and characterise the receptivity of the baseflow introduced in the next section.

\section{Baseflow computation}

\begin{figure}[t]
    \includegraphics[width=\linewidth]{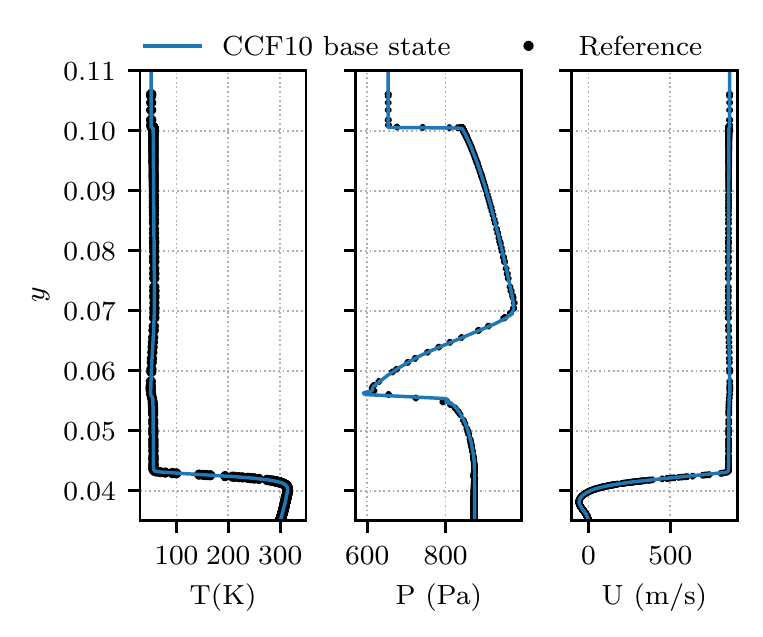}
    \caption{Validation of the base-flow by comparison with a reference solution at the cylinder-flare junction \cite{paredesBoundaryLayerInstabilitiesCone2022}.}
    \label{fig:ccf12-valid}
\end{figure}
\begin{figure*}[t]
    \centering{
    \begin{subfigure}[t]{0.45\textwidth}
        \centering
        \includegraphics[height=0.8\textwidth]{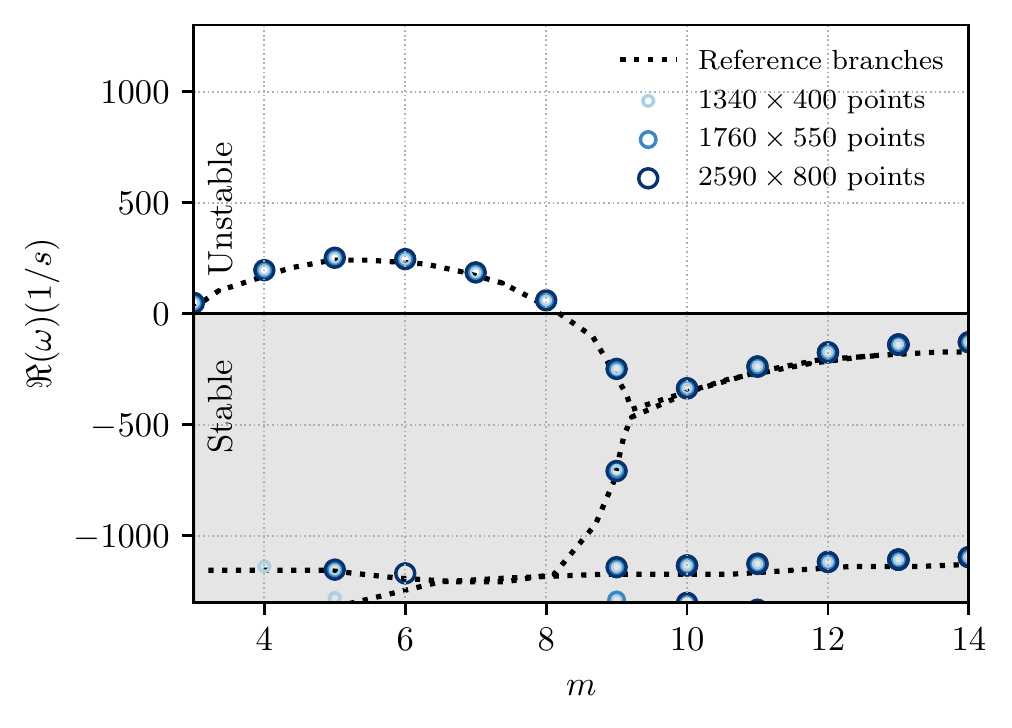}
        \caption{$R_n=0.1$mm}
        \label{fig:valid_global_a}
    \end{subfigure}
    \hspace{1cm}
    \begin{subfigure}[t]{0.45\textwidth}
        \centering
        \includegraphics[height=0.8\textwidth]{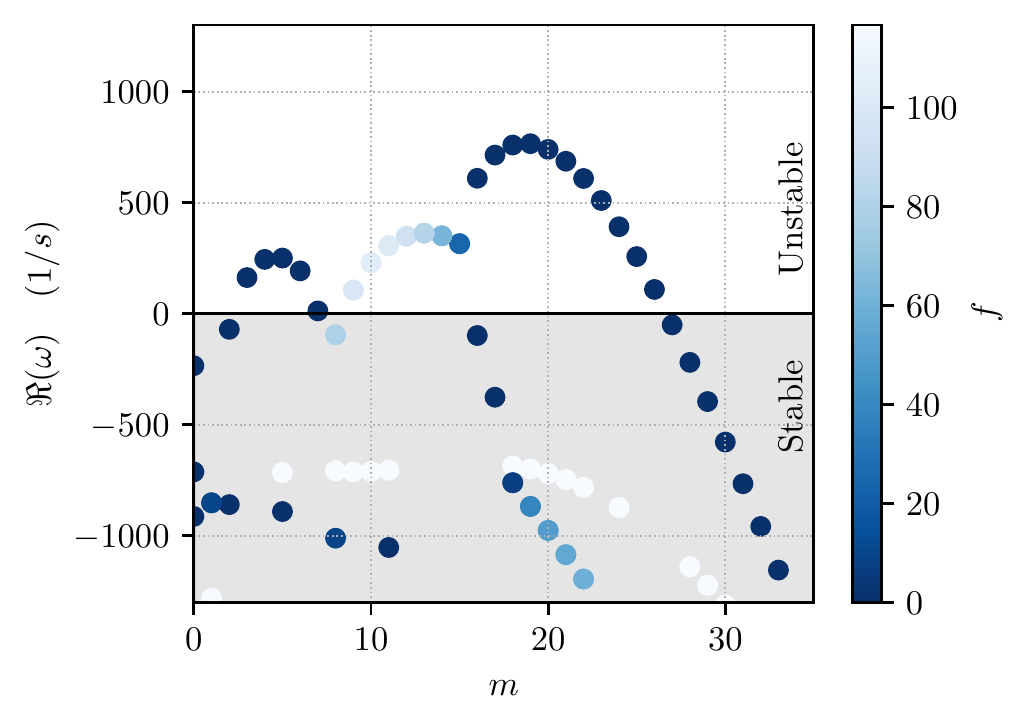}
        \caption{$R_n=5$mm}
        \label{fig:valid_global_b}
    \end{subfigure}}
    \caption{Eigen values growth-rates as a function of $m$ for both nose radius. The left spectrum shows a grid convergence study. The right spectrum eigenvalues are coloured by their frequencies (Hz)}
    \label{fig:valid_global}
\end{figure*}
The laminar flow around the CCF geometry comprises a detached shock, expansion waves, separation and reattachment shocks around the bubble. These features poses a challenge to the computation of fixed-points for different Reynolds numbers. This section outlines the flow conditions and details the procedure used to compute fixed points. Finally the baseflows are described and validated.

The flow around CCF10 is studied at a Mach number $M_\infty=6.0$ for a unit Reynolds number of $\text{Re}=11.5\times10^6$. The domain extends from the wall to the free-stream and from the stagnation point to the end of the flare. An isothermal wall at $T_w=300K$ is considered. These conditions relate to experiments performed in quiet and noisy wind tunnels \cite{benitezInstabilityMeasurementsAxisymmetric2020} and are summarised in Table \ref{tab:conditions}
\begin{table}[]
    \centering
    \begin{tabular}{l c c c c }
        Case & Re & $R_n$ & $T_\infty$ & $\rho_\infty$  \\ \hline
         R-$R_n$ & $m^{-1}$ & $mm$ & $K$ & $kg/m^3$  \\ \hline
        R01 & $11.5\times10^6$ & 0.1 & $51.4$K & $4.43\times10^{-2}$ \\
        R5 & $11.5\times10^6$ & 5 & $51.4$K & $4.43\times10^{-2}$  \\
    \end{tabular}
    \caption{Flow conditions considered}
    \label{tab:conditions}
\end{table}

The computation of a fixed point uses the following steps : first, a preliminary flow is computed on a coarse structured mesh which is not shock-aligned. Second, once this coarse solution reaches a sufficient convergence level, the shock boundary is extracted and a fine shock-aligned grid is generated. The coarse solution is linearly interpolated on the fine grid and serves as the initial guess to a pseudo transient continuation technique \cite{crivelliniImplicitMatrixfreeDiscontinuous2011}. The solution vector update $\pmb q^{n+1} = \pmb q^{n} + \delta \pmb q^n$ is given by the computation of the step $\delta q$ as,
\begin{align}
    \left( \frac{\mathbf I}{\Delta t} + \mathbf{L}\pmb q^n \right) \delta q^n = - \mathbf{R}(\pmb q^n). \label{eq:newton}
\end{align}

The residuals $\mathbf{R}(\pmb q^n)$ are computed using a 7-th order FE-MUSCL scheme with a high order dissipation and Jameson-like shock capturing controlled by a shock sensor, resulting in low dissipation levels \cite{sciacovelliAssessmentHighorderShockcapturing2021}. The jacobian is computed with the same numerical methods through algorithmic differentiation. Between 5 and 20 pseudo-Newton iterations are required to converge the fixed-point to a normalised residual value of $||\mathbf{R}(\pmb q_0)||_2 \leq 10^{-14}$. 

Baseflows $\pmb q_0$ are shown in Fig. \ref{fig:ccf12-bf} for cases R01 and R5. Observing the flow features for $R_n=0.1$mm \& $R_n=5$mm reveals two main differences. First, starting from the nosetip, the R5 case shows a thicker entropy layer up to the cone-cylinder junction induced by the detached shock. Whereas for the R01 case, the entropy layer remains thin and is actually following closely the boundary layer height. The entropy layer effects are known to affect the growth of instabilities on hypersonic blunt bodies \cite{stetsonNosetipBluntnessEffects1983} and will be investigated in the coming sections. The second noticeable difference between the two cases is the length of the separation bubble, increasing the bluntness leads to a bigger separation length induced by the reduced edge Mach number $M_e$ and the thicker boundary layer.

The computed fixed point for $R_n = 0.1mm$ is validated against the previous study of \cite{paredesBoundaryLayerInstabilitiesCone2022} for the same conditions but different numerical methods. Fig. \ref{fig:ccf12-valid} shows a perfect agreement between the two solutions hence validating both computations of the hypersonic laminar baseflow.


For each fixed-points, the jacobian is computed and is used to probe the linear stability of this two flows.

\section{Global stability}

This section aims at solving the eigenvalue problem introduced in Eq. \ref{eq:EVP} for successive discrete azimuthal wavenumbers $m$. For case R01 some modes of the spectrum are compared and validated against the previous findings of \cite{paredesBoundaryLayerInstabilitiesCone2022}. On the other hand the discussion on the global stability case R5 constitutes a new set of results.

Each spectrum is obtained from the jacobian using an iterative LU+Arnoldi algorithm as depicted in \cite{poulainBROADCASTHighorderCompressible2023}, the computations uses the PETSc/SLEPc routines along with the MUMPS library. This iterative solver is set to converge the 5 eigenvalues closest to a target $\tau = 1000.0 + 0.0i$ using a shift-invert strategy. As the unstable modes and spectrum computation are dependent on the baseflow discretisation, a spectrum convergence is carried out beforehand. The Fig. \ref{fig:valid_global_a} displays this convergence in the wavenumber-growth-rate frame. A reference computation from \cite{paredesBoundaryLayerInstabilitiesCone2022} is used to plot the modes branches. Note that since the domain is axisymmetric, only discrete values of $m$ can physically be allowed to exist. For the spectrum convergence, three grids with increasing number of streamwise and vertical points are considered. All of them yield consistent results with the reference, therefore confirming that even a coarse grid resolution allows to properly converge these global structures. Although, as it will be required later for the forced analysis, a grid of $N_x\times N_y = 4030 \times 550$ will be used for the subsequent computations.  

Aside from the validation analysis, comparing the Figs. \ref{fig:valid_global_a} \&  \ref{fig:valid_global_b} unveils significant differences between the global dynamics of the bubble between $R_n=0.1$mm and $R_n=5$mm. The spectrum of the blunt configuration reveals three branches of unstable modes. All the eigen values are coloured by their frequency to distinguish unsteady dynamics. Looking at the spectrum of Fig. \ref{fig:valid_global_b} reveals two steady branches for $m\in[3,7]$ and $m\in[15,26]$. This latter branch dominates the growth-rates of the spectrum. A third unsteady branch is visible for $m\in[8,14]$, these modes have a frequency between $f=60$Hz and $f=90$Hz, suggesting quasi-steady dynamics in comparison to the flow characteristic time scales.

\begin{figure}[tb]
    \centering
    \includegraphics[width=0.45\textwidth]{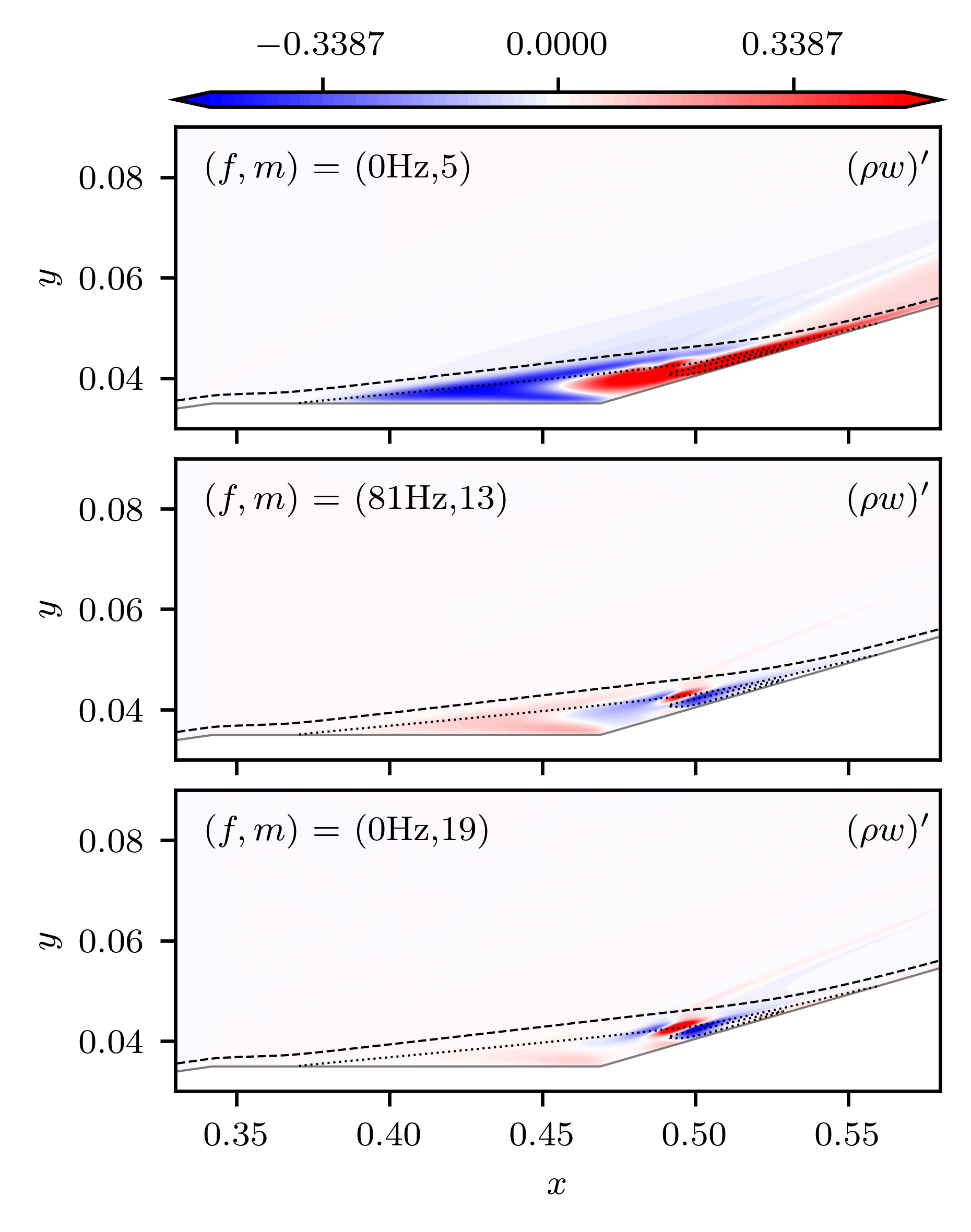}
    \caption{Eigenfunctions of the leading global modes for $R_n=5$mm. Normalised contours of disturbances transverse momentum ; (\dashline) boundary layer ; (\dotline) separated region.}
    \label{fig:glob_modes}
\end{figure}

All the unstable modes arising in Fig. \ref{fig:valid_global} appear to be structures modulating the bubble shape with wavenumbers $m>0$. Therefore, the growth of these structures leads to a three-dimensionalisation of the baseflow. Similar mechanisms were found by \cite{gsOnsetThreedimensionalitySupersonic2018,caoTransitionTurbulenceHypersonic2022,lugrinMultiscaleStudyTransitional2022}. The eigenfunctions of the dominating modes for each unstable branch of Fig. \ref{fig:valid_global_b} are plotted in Fig. \ref{fig:glob_modes}. The mode at $(f,m)=(0,5)$ is representative of bubble breathing mode. Its structure is similar for $R_n=0.1$mm and $R_n=5$mm, suggesting that it comes from the same underlying physical mechanism in both cases. The eigenfunction for $(f,m)=(81,13)$ seem to be mainly amplified near the reattachment point, displaying a strong transverse motion at this position. Finally, the instability for $(f,m)=(0,19)$ also seem to be a reattachment mode and show a transverse motion very similar to the mode $(f,m)=(60,13)$. Further analysis have to be made with the eigenfunctions in order understand the physical origins of these similarities.   

%

The global dynamics obtained from the linear stability analysis of Eq. \ref{eq:EVP} revealed the presence of multiple steady and unsteady structures susceptible to alter the laminar base flow and even lead to turbulence. In the non-linear regime, these global mode will saturate and may interact with the convective instabilities originating from the external disturbances, hence, the next section aims at characterising these convective instabilities. 

\section{Optimal forcing}

The response of the baseflow to external disturbances is analysed using Eq. \ref{eq:resolvent-evp}. Maps of the optimal gain in $(\omega, m)$ are computed for each case and allow to see the most amplified mechanism at a given frequency and wavenumber along the object for both nose radii. 

\begin{figure*}[htb]
    \centering
    \begin{subfigure}[t]{0.9\textwidth}
        \centering
        \includegraphics[width=\textwidth]{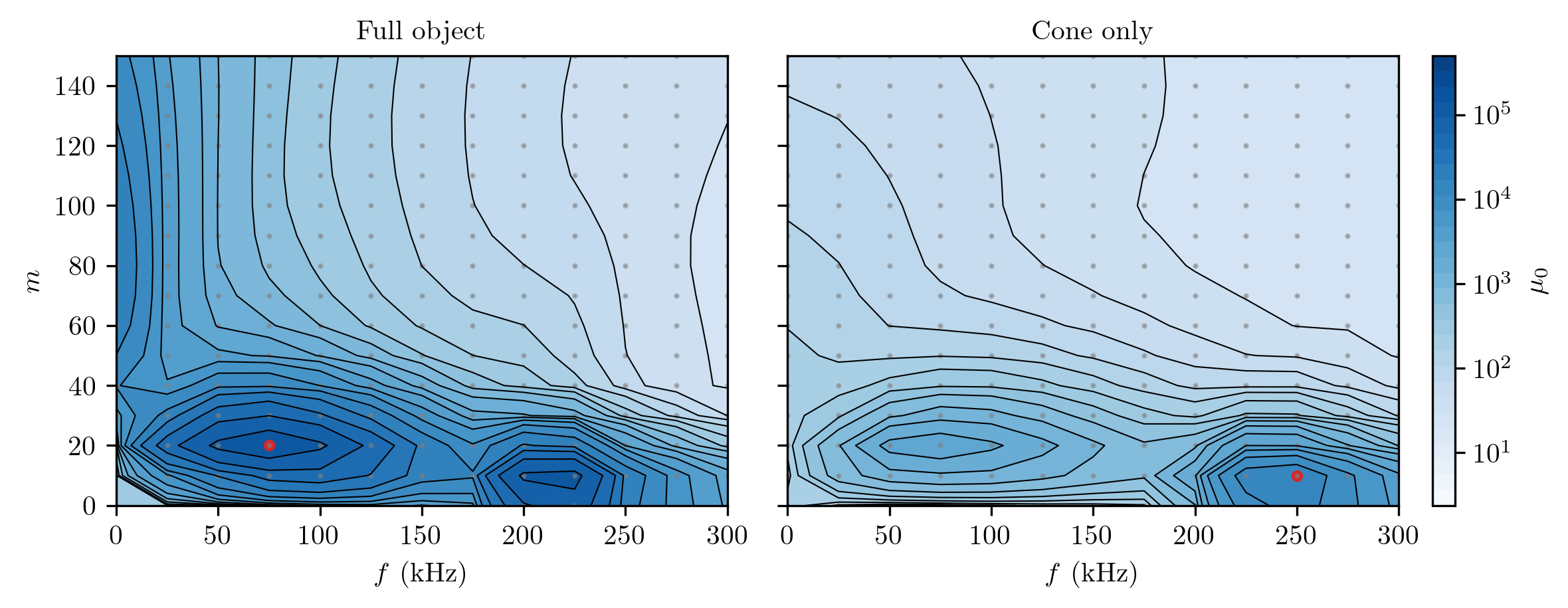}
        \caption{$R_n=0.1$mm}
        \label{fig:gains-map-R01_a}
    \end{subfigure}
    \\
    \begin{subfigure}[t]{0.9\textwidth}
        \centering
        \includegraphics[width=\textwidth]{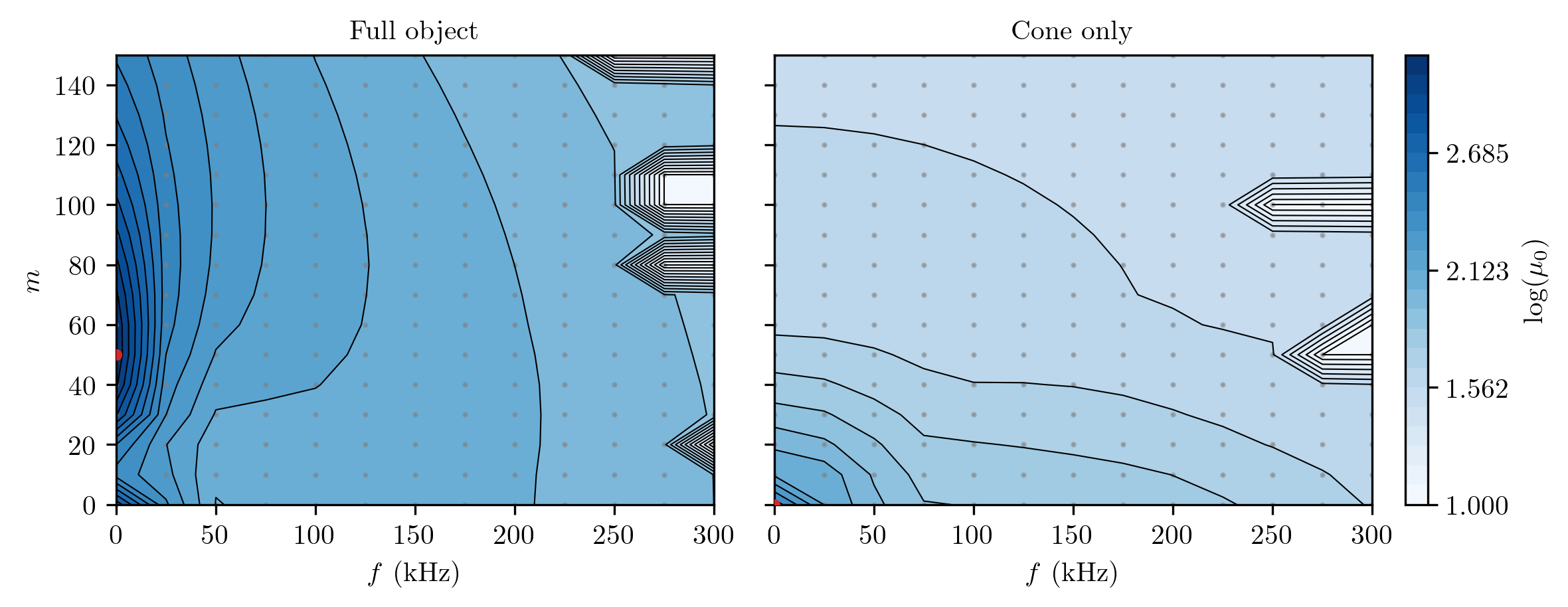}
        \caption{$R_n=5$mm}
        \label{fig:gains-map-R01_b}
    \end{subfigure}
    \caption{Maps of optimal gain for the R01 case for the full geometry (left) and the cone only (right). Dominating peaks :  (\textcolor{mplred}{$\bullet$}). Light grey dots show the points actually computed for building the maps.}
    \label{fig:gains-map-R01}
\end{figure*}
\begin{figure}[htb]
    \centering
    \begin{subfigure}[t]{0.5\textwidth}
        \centering
        \includegraphics[width=\textwidth]{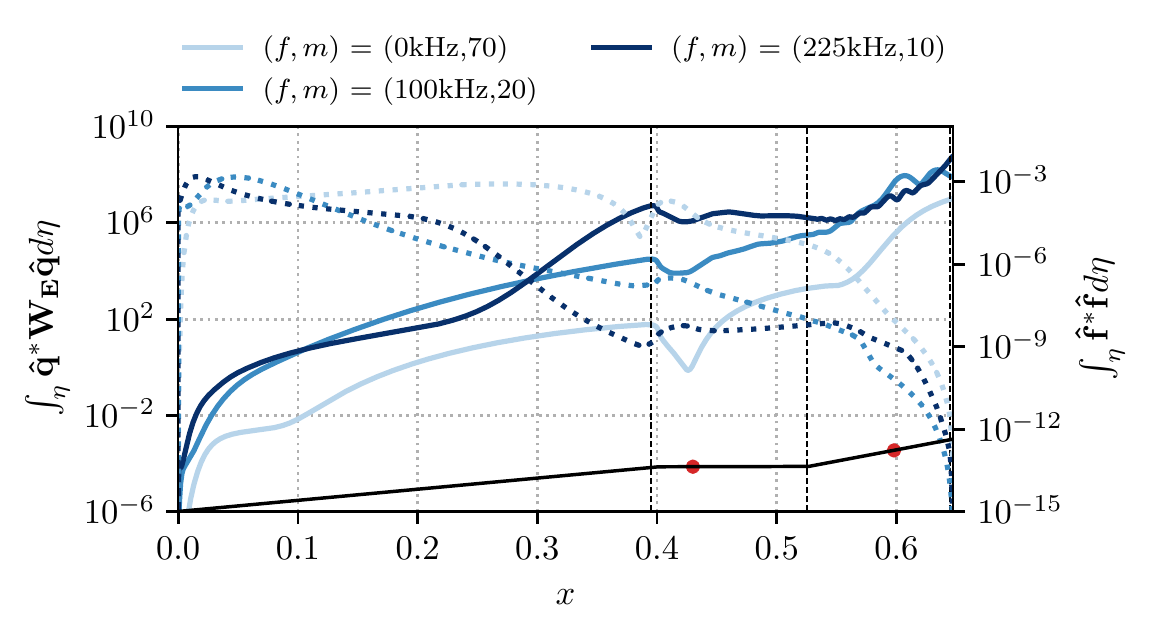}
        \caption{$R_n=0.1$mm}
        \label{fig:gains_a}
    \end{subfigure}
    \\
    \begin{subfigure}[t]{0.5\textwidth}
        \centering
        \includegraphics[width=\textwidth]{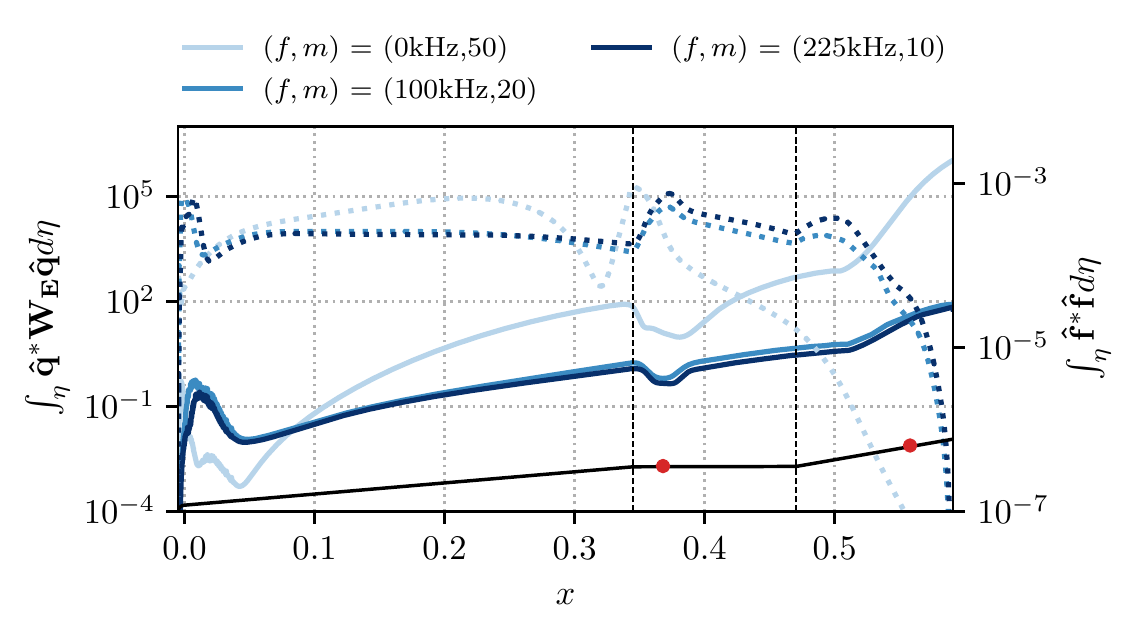}
        \caption{$R_n=5$mm}
        \label{fig:gains_b}
    \end{subfigure}
    \caption{Integrated energy of optimal responses $\hat{\pmb q}$ (\contline) and forcings $\hat{\pmb f}$ (\dotline) for the different peaks observed in Figs. \ref{fig:gains-map-R01_a} \& \ref{fig:gains-map-R01_b}. \textcolor{mplred}{$\bullet$} : separation and reattachment points. In the legend from left to right : streaks, first mode, second mode, frequencies are in kHz. The object shape is depicted at the bottom for clarity.}
    \label{fig:gains}
\end{figure}
\begin{figure}[tb]
    \centering
    \includegraphics[width=0.48\textwidth]{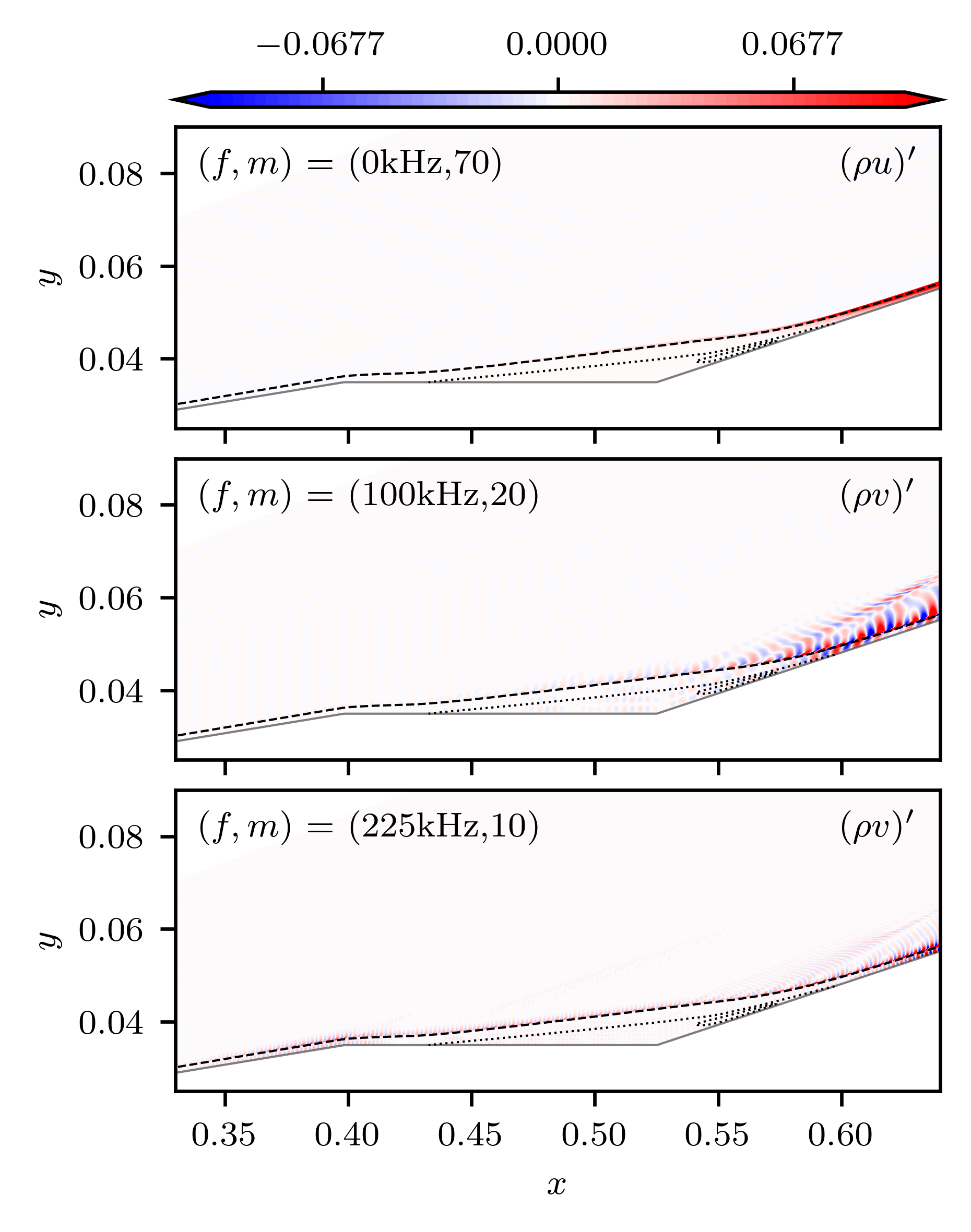}
    \caption{Optimal responses corresponding to the leading resolvent modes for $R_n=0.1$mm. Normalised contours of disturbances momentum, saturated at 10\% of the normalised value ; (\dashline) boundary layer ; (\dotline) separated region.}
    \label{fig:r01_modes}
\end{figure}

In order to quantify the effects of the separation region, these maps are computed for both the full geometry and a subdomain restricted to the cone. Indeed, as the baseflow can be globally unstable on the cylinder-flare portion, an input-output analysis on the steady-homogenous fixed point $\pmb q_0$ computed through Eq. \ref{eq:newton} could be ill-posed as the system bifurcated. Hence, solving Eq. \ref{eq:resolvent-evp} for the cone region, which is globally stable and independent of the bifurcated state due to the supersonic regime, allows for a consistent optimal response computation and gives a linear estimation of the wave content that might enter the bubble in a non-linear scenario. 

Figure \ref{fig:gains-map-R01} regroups the gain maps for cases R01 and R5. First, a baseline analysis is made for the R01 case before showing the variations induced by the blunt nose of case R5. Figure \ref{fig:gains-map-R01_a} shows the optimal gains map for both the full geometry and the region restricted to the cone only. In summary, three main resonance regions appear : the dominating resonance peak is found near $(f,m)=(75\text{kHz}, 20)$ and seems to be related to the family of first mode instabilities. Next, a resonance is visible for $(f,m)=(225\text{kHz}, 10)$ and is experimentally verified to be related to the family of second-mode instabilities. Finally, a third resonance peak is found for steady waves at $(f,m)=(0\text{kHz},70)$ identified as streamwise streaks. 

Looking at the cone-only gain map highlights different dynamics, as the dominating peak is now the second mode at $(f,m)=(250\text{kHz}, 10)$, which is consistent with previous literature results on canonical sharp cone boundary layers at $M_\infty=6.0$. The first mode broadband peak remains present for $(f,m)=([50,100]\text{kHz}, 20)$ but shows an amplification at least one order of magnitude smaller that the second mode. Finally, the main difference with a resolvent computation on the full geometry resides in the absence of a clear streaks amplification peak at the previously observed $(f,m)=(0\text{kHz},70)$ region. This does not necessarily imply the total absence of streaks as these can still have a marginal amplification from non-normality on the cone $(\mu_0 \approx 10^2)$. Additionally, the maximum value of $m$ chosen for the gain map might not be high enough for capturing the higher wavenumber streaks that might appear close to the nosetip where the boundary layer height is smaller. 

Comparing the previous results with the gain maps of the R5 case in Fig. \ref{fig:gains-map-R01_b} shows that the bluntness induces substantial differences in the baseflow response to an external forcing. Looking at the full object gain map, the peaks of the first and second mode instabilities observed for case R01 have vanished. The R5 baseflow displays a weak non-normal amplification for frequencies above $f=50$kHz. The only observed peak is related to the steady streaks having a maximum gain : $\mu_0=1.4\times 10^3$ at $(f,m)=(0,50)$. The gain map restricted to the blunt cone of case R5 shows the same absence of strongly amplified peaks for the frequencies and wavenumbers corresponding to the first and second modes found in Fig. \ref{fig:gains-map-R01_a}. 

As such, these latter frequency-wavenumber ranges can no-longer be related to first and second mode waves as no clear resonance peak emerges. Instead, the gain map region for $f\in[25,300]$kHz and $m\in[0,150]$ is most likely related to weak non-normal amplification mechanisms. This hypothesis has to be further investigated in following studies by looking at the rank of the resolvent operator in this region. Nonetheless, the behaviour of the waves at $(f,m)=(75\text{kHz}, 20)$ and $(f,m)=(225\text{kHz}, 10)$ will also be studied for case R5 in order to have a comparison basis between the sharp and blunt dynamics.

These differences in the optimal gains maps between the sharp and blunt nosetips can be related to the transition reversal phenomenon encountered for hypersonic boundary layers \cite{stetsonNosetipBluntnessEffects1983}. For the blunt nose, the entropy layer added to the decrease in edge Mach number induced by the strong shock might suppress the support of the unstable waves found for R01. It should be precised that the gain maps for case R5 contains some unresolved gains value (white regions), these arise from difficulties in solving Eq. \ref{eq:resolvent-evp} in presence of the strong detached shock. The discontinuity also induces a gain peak at $(f,m)=(0,0)$. This gain peak can not be directly interpreted as a convective instability and seem to be related to an infinitesimal shock motion induced by the forcing of the momentum equations which reproduce the effect of infinitesimal variations in the Mach number. 

%
\begin{figure*}[tb]
    \centering
    \includegraphics[width=\textwidth]{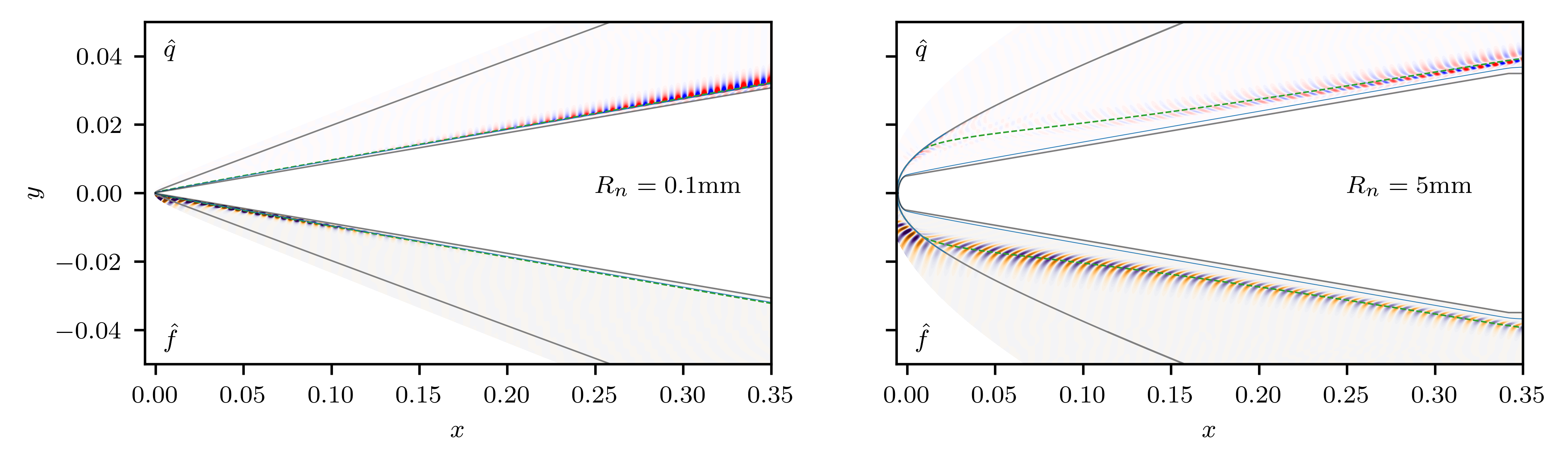}
    \caption{Comparing optimal responses $\hat{\pmb q}$ and optimal forcings $\hat{\pmb f}$ at $(f,m)=(100\text{kHz},20)$ for cases R01 (left) and (R5). Normalised contours of $(\rho v)'$ ; (\cdashedline{mplgreen}) entropy layer ; (\ccontline{mplblue}) boundary layer edge $(h_t/h_{t_\infty} = 0.995)$ ; (\contline) detached shock.}
    \label{fig:compare_res}
\end{figure*}

Further insights can be gained by looking at the spatial growth along the object of the structures corresponding to the three peaks discussed above. In supplement to the global information provided by the gain maps, it helps to understand how the optimal forcings and responses evolve around the different regions. Considering the mesh to be quasi-wall normal, up to a small approximation error, the integrated energy of the modes is computed in the wall-normal grid direction $\eta$ as,
\begin{align}
    E_y = \int_\eta \hat{\pmb q}^* \mathbf{W}_E \hat{\pmb q} \text{d}\eta.
\end{align}
Where, $\hat{\pmb q}$ and $\mathbf{W}_E$ represents the resolvent response mode and its associated energy weighting matrix as stated in Sec. \ref{sec:lin-framework}. A similar procedure holds for the optimal forcings as well. 

Figure \ref{fig:gains} displays the integrated energy of the optimal responses in thick lines and the associated forcing norm in dotted lines. For case R01 in Fig. \ref{fig:gains_a}, it can be immediately noticed that the overall energy of the first and second modes remains two or three order of magnitude higher than the steady streaks at $m=70$ along the whole geometry. These former two modes displays similar energy levels on the cone up to $x\approx0.28$m. From that point, the $(100,20)$ wave shows a steady amplification, whereas the $(225,10)$ wave exhibits an increased growth that leads to its dominance at the end of the cone. This increased growth of the second mode wave is explained by the dependency of this family of instabilities to the boundary layer height. For case R5, the same modes are plotted for unsteady waves as a comparison basis. These waves undergo a weak amplification consistent with the low gain observed in Fig. \ref{fig:gains-map-R01_b}. Further discussions on the differences of these non-normal mechanisms for R01 and R5 will be provided in a subsequent section. 

A second point of interest is the expansion fan of the cone-cylinder junction after $x=0.4$, it displays a stabilising effect for the three considered waves in case R01 and R5, especially for the streaks. Oppositely, the expansion fan is a favourable region for receptivity as it can be seen by the sharp and localised increase in the forcings norm. Next, the instabilities are convected through the separated flow. For this region, the curves of Fig. \ref{fig:gains_a} are supplemented by the corresponding optimal response eigenfunctions of case R01 shown in Fig. \ref{fig:r01_modes}. Focusing on Fig. \ref{fig:gains_a}, along the cylinder, the first-mode and the streaks instabilities remain amplified, especially the streaks which shows a sudden growth rate increase at the onset of separation. On the contrary, the second mode waves get slightly damped, suggesting a stabilising effect of the separated region on these convective instabilities. Near the reattachment, all modes display an increase in energy, most probably linked with the strong re-compression, and keep amplifying along the flare. Especially, the first mode mode wave has an energetic content at the end of the flare similar to the second mode wave. This states that both mode have to be equally considered when studying a transition scenario on this geometry at similar conditions. The response modes shown in Fig. \ref{fig:r01_modes} indicate that the spatial support of the streaks is mainly located within the boundary layer and after the separation region. On the other hand both first and second mode waves show some fluctuations along and under the shear layer of the bubble around $x=0.52$m. 

Focusing on the forcings location along the full object, some comments can be made on the receptivity support of these instabilities. For case R01, the first and second mode forcings shown in Fig. \ref{fig:gains_a} indicate that the region close to the nosetip is the most sensitive to the receptivity process. After $x=0.1$m, the forcings energy steadily decreases, offering lesser projection support for external disturbances. A similar behaviour is observed for the streaks forcing vector, although, its energy remains quite constant up to the end of the cone, suggesting that the disturbances leading to the streaks on the flare are actually growing all along the geometry, a similar observation is made for the streaks of case R5. 

In relation to the forcing spatial support, additional information in the difference of optimal gains between the sharp and blunt case can be gained by looking at the resolvent optimal eigenfunction around the nosetip. Figure \ref{fig:compare_res} provides this comparison for the wave at $(f,m)=(100\text{kHz}, 20)$. This frequency and eigenvalue corresponds to the peak of the first mode instability for case R01 and to a weak and broadband non-normal amplification mechanism for case R5. For both geometries, the entropy layer, the boundary layer and the detached shock-wave of the respective baseflows are plotted in order to illustrate the differences in spatial support of the non-normal mechanisms. Consistently with previous literature results, for case R01, the optimal forcings are located upstream on the cone, close to the nosetip and within the boundary layer. The associated response is amplifying downstream around the boundary layer edge. For the wave of R5 at the same frequency and wavenumber, different spatial supports are observed. The forcing is no longer restrained close to the nosetip and spread along the cone. Moreover, its most energetic region is mainly located within the entropy layer and does not attain the boundary layer. The associated response is located downstream on the cone but is also mainly amplifying along the entropy layer edge and does not seem to fluctuate within the boundary layer. 

Such differences support the hypothesis of a different mechanism at the origin of the wave amplification on the blunt cone. Similar entropy layer modes were found by previous DNS studies for moderately blunt cones \cite{hartmanNonlinearTransitionMechanism2021,paredesMechanismFrustumTransition2020}.

\begin{figure}[htb]
    \centering
    \includegraphics[width=\linewidth]{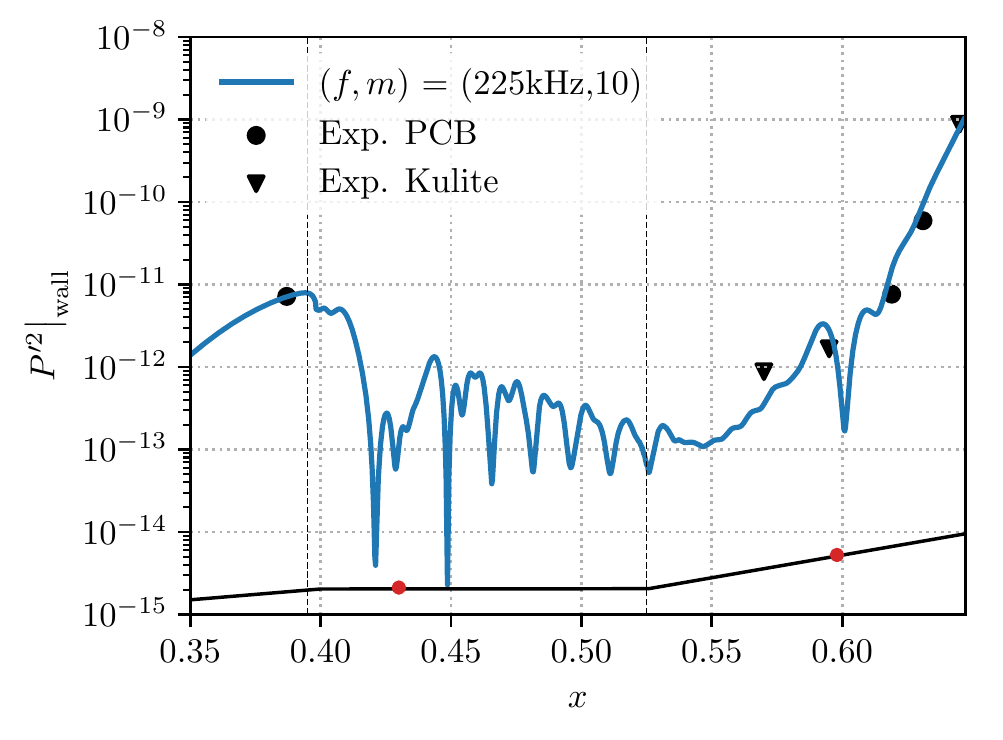}
    \caption{Comparison of the R01 case second-mode pressure response at the wall with wind tunnel measurements from unsteady pressure sensors for the same conditions \cite{paredesBoundaryLayerInstabilitiesCone2022}.}
    \label{fig:bcast-expe}
\end{figure}
    Finally, an assessment of the resolvent ability to predict hypersonic instabilities growth around the CCF10 object is made in Fig. \ref{fig:bcast-expe} by comparing the optimal second-mode $(f,m)=(225\text{kHz},10)$ pressure at the wall, to experimental results \cite{paredesBoundaryLayerInstabilitiesCone2022}. The computation is shown along with the spectra peak value at $f\approx225$kHz acquired during a wind tunnel test at the same free-stream conditions. These spectra were obtained with wall mounted PCB and Kulite pressure sensors placed at different stations at the end of the cone and along the flare. Since the computation is linear, an initial amplitude $A_0$ has to be set for the resolvent mode to have a physically meaningful amplitude, this value is calibrated to match the first PCB sensor energy at $x=0.387$m, leading to $A_0 = 1.2 \times 10^{-20}$. The computation and the sensors wall pressure shows a close agreement in the reattachment region, the resolvent pressure amplitude function is matching the experimentally observed trends in pressure fluctuations. Such a result confirms the ability of the current linear framework to predict accurate hypersonic instabilities evolution in presence of complex laminar flow features.

\section{Conclusion}

The present study provides a comprehensive review of an ongoing numerical investigation aiming at uncovering the linear instabilities of the hypersonic laminar flow around the CCF10 geometry. This research works towards providing a deeper understanding of the transition to turbulence under experimental and flight conditions by mapping the various linear instabilities at play.

To achieve this objective, the numerical framework was first validated against previous results, followed by a mapping and comparison of the linear mechanisms on both sharp and blunt configurations. The validation procedure demonstrated the accuracy of the BROADCAST toolbox in providing fixed points as a starting point for the global stability investigation. Additionally, the obtained global modes for the sharp case were found to compare well with the literature's spectrum for the same conditions.

Further analysis revealed differences in the laminar bubble dynamics for the sharp and blunt global spectra, with an increased number of unstable branches and the presence of unsteady bubble modes for the R5 case. The origin of these different dynamics is yet to be explained and may be investigated through the structural sensitivity of the baseflow.

The study also highlighted the differences in amplifying mechanisms between the sharp and blunt configurations. The sharp configuration displayed clear amplification peaks for the first and second Mack modes, whereas the blunt configuration only showed an amplification peak for steady streaks.

Investigating further these entropy layer modes allowed to observe their associated optimal forcings structures. Such forcing waves were scarcely discussed in the literature and provide a good starting point for explaining the origin of the observed entropy layer instabilities observed in the aforementioned DNS studies.

Finally, a direct comparison with experimental data obtained from the Purdue BAM6QT quiet wind tunnel confirmed the framework ability to accurately estimate convective instabilities evolution along the object. 

Overall, this preliminary study validated the stability framework and provided insights into the linear dynamics around the Cone-Cylinder-Flare geometry. Future work will focus on explaining the physical mechanisms at the origin of the discrepancies observed between the sharp and blunt cone dynamics. This set of results will then help to better understand future non-linear simulations and experimental campaigns.

\bibliographystyle{plain}
\bibliography{references}

\end{document}